\documentclass[twocolumn,aps,prb,groupedaddress,showpacs,floatfix]{revtex4-1}
\usepackage{graphicx}
\usepackage{amsmath}

\begin{document}

\title{Silicon spin chains at finite temperature: dynamics of Si(553)-Au}
\author{Steven C. Erwin}
\email{steve.erwin@nrl.navy.mil}
\affiliation{Center for Computational Materials Science, Naval Research Laboratory, Washington, DC 20375}
\author{P.C. Snijders}
\affiliation{Materials Science and Technology Division, Oak Ridge National Laboratory, Oak Ridge, TN 37830}
\date{\today}

\begin{abstract}
  When gold is deposited on Si(553), the surface self-assembles to
  form a periodic array of steps with nearly perfect structural order.
  In scanning tunneling microscopy these steps resemble
  quasi-one-dimensional atomic chains.  At temperatures below $\sim$50
  K the chains develop a tripled periodicity.  We recently predicted,
  on the basis of density-functional theory calculations at $T=0$,
  that this tripled periodicity arises from the complete polarization
  of the electron spin on every third silicon atom along the step; in
  the ground state these linear chains of silicon spins are
  antiferromagnetically ordered. Here we explore, using ab-initio
  molecular dynamics and kinetic Monte Carlo simulations, the behavior
  of silicon spin chains on Si(553)-Au at finite temperature.
  Thermodynamic phase transitions at $T>0$ in one-dimensional systems
  are prohibited by the Mermin-Wagner theorem. Nevertheless we find 
  that a surprisingly sharp onset occurs upon cooling---at
  about 30 K for perfect surfaces and at higher temperature for
  surfaces with defects---to a well-ordered phase with tripled
  periodicity, in good agreement with experiment.
\end{abstract}

\maketitle

\section{Introduction}

Linear atomic chains of metal atoms on semiconductor surfaces offer,
in principle, the physical realization of phenomena predicted
theoretically for one-dimensional model systems. In practice, however,
unanticipated interactions can often complicate the picture and lead
to behavior not easily explained by simple models.
In this article we demonstrate theoretically how the complex interactions
among polarized electron spins in silicon surface states determine the
observed behavior of a well-studied atomic chain system, Si(553)-Au,
over a wide range of temperatures. The methods developed here and the
resulting predictions---which are qualitatively and quantitatively
consistent with experimental observations---are also likely to apply
more broadly to other vicinal Si/Au chain systems, such as Si(557)-Au.

The Si(553)-Au surface was first investigated in
Ref.~\onlinecite{Crain2003}, an experimental study which established
that the electronic band dispersion and fermi surface were indeed
those of a nearly one-dimensional metal. Since then, numerous lines of
research have emerged.  Efforts to determine the
basic atomic structure of the surface have been based on data from
diffraction experiments\cite{Ghose2005,Takayama2009,Voegeli2010} and
on the results of theoretical total-energy
calculations.\cite{Riikonen2005,Riikonen2006,Riikonen2008,Krawiec2010}
These were greatly aided by the first definitive determination of the
coverage of Au atoms on Si(553)-Au.\cite{Barke2009} Other
investigations have explored the properties of finite-length
chains\cite{Crain2005,Crain2006} as well as various native
defects\cite{Okino2007a} and foreign adsorbates
\cite{Ryang2007,Okino2007,Ahn2008,Kang2009,Kang2009,Kang2009b,Nita2011,Krawiec2013}
on the nominally clean Si(553)-Au surface.

One particularly interesting line of research has focused on the
collective behavior in Si(553)-Au that emerges at low temperature.
Ideal one-dimensional metals with partially filled bands exhibit a
broken symmetry at low temperature, namely a charge-density wave
arising from the Peierls instability. Indeed, broken symmetries
in Si(553)-Au were observed using scanning tunneling microscopy
(STM)  in Refs.\ \onlinecite{Ahn2005} and
\onlinecite{Snijders2006}. Images acquired at room temperature showed alternating bright
and dim rows with unit periodicity $a_0$ along the rows.  Below
$\sim$50 K these rows separately developed higher-order periodicity: a
tripled period (3$a_0$) along the bright rows and a doubled period
(2$a_0$) along the dim rows. Subsequent review articles have discussed
possible explanations for these
observations.\cite{Snijders2010,Hasegawa2010}

Notwithstanding the fact that Peierls instabilities lead to
higher-order periodicity, a completely different theoretical
explanation for the coexisting triple and double periodicities in
Si(553)-Au was proposed in Ref.\ \onlinecite{Erwin2010}. The key idea,
which was based on the results of density-functional theory (DFT)
calculations, was that the ground state of Si(553)-Au is spin
polarized. In the DFT ground state, the silicon atoms that
comprise the steps on this vicinal surface have dangling bonds, every
third of which is occupied by a single fully polarized electron while
the other two are doubly occupied.  The bright rows seen in empty-state
STM images arise from these step-edge silicon atoms.  At low
temperature the $3a_0$ peaks that appear in this row are from the
spin-polarized atoms, which have precisely this periodicity.  The DFT
ground state also reveals a period doubling within the row of Au atoms.
Both of these higher-order periodicities disappear if spin
polarization is suppressed in the calculation, providing compelling
evidence that spin polarization is the primary mechanism underlying the
observed symmetry breaking in Si(553)-Au.

Experiments were subsequently carried out to look for a spectroscopic
signature of this predicted spin-polarized ground state. The DFT
calculations showed that an unoccupied state should exist 0.5 eV above
the fermi level and be localized at the polarized silicon
atoms.\cite{Erwin2010} The existence and spectral and spatial location
of this state was indeed confirmed by two-photon photoemission
\cite{Biedermann2012} and by scanning tunneling
spectroscopy.\cite{Snijders2012}

The predictions of Ref.\ \onlinecite{Erwin2010} only addressed the
zero-temperature ground state of Si(553)-Au. Left unanswered in that
work was the question of how the broken-symmetry ground state evolves
to have normal $a_0$ periodicity above $\sim$50 K. This article
addresses that question from a theoretical and computational
perspective. Although it may seem obvious that thermal fluctuations
are important, the nature of these fluctuations turns out to be
unexpectedly subtle. Nevertheless, we derive here a number of detailed
qualitative as well as quantitative predictions that can easily be
tested experimentally. The results of these tests will furnish
additional evidence for evaluating the validity of the basic mechanism
proposed in Ref.\ \onlinecite{Erwin2010}.

\section{Ground state configuration}

\begin{figure}[tb]
\includegraphics[width=8cm]{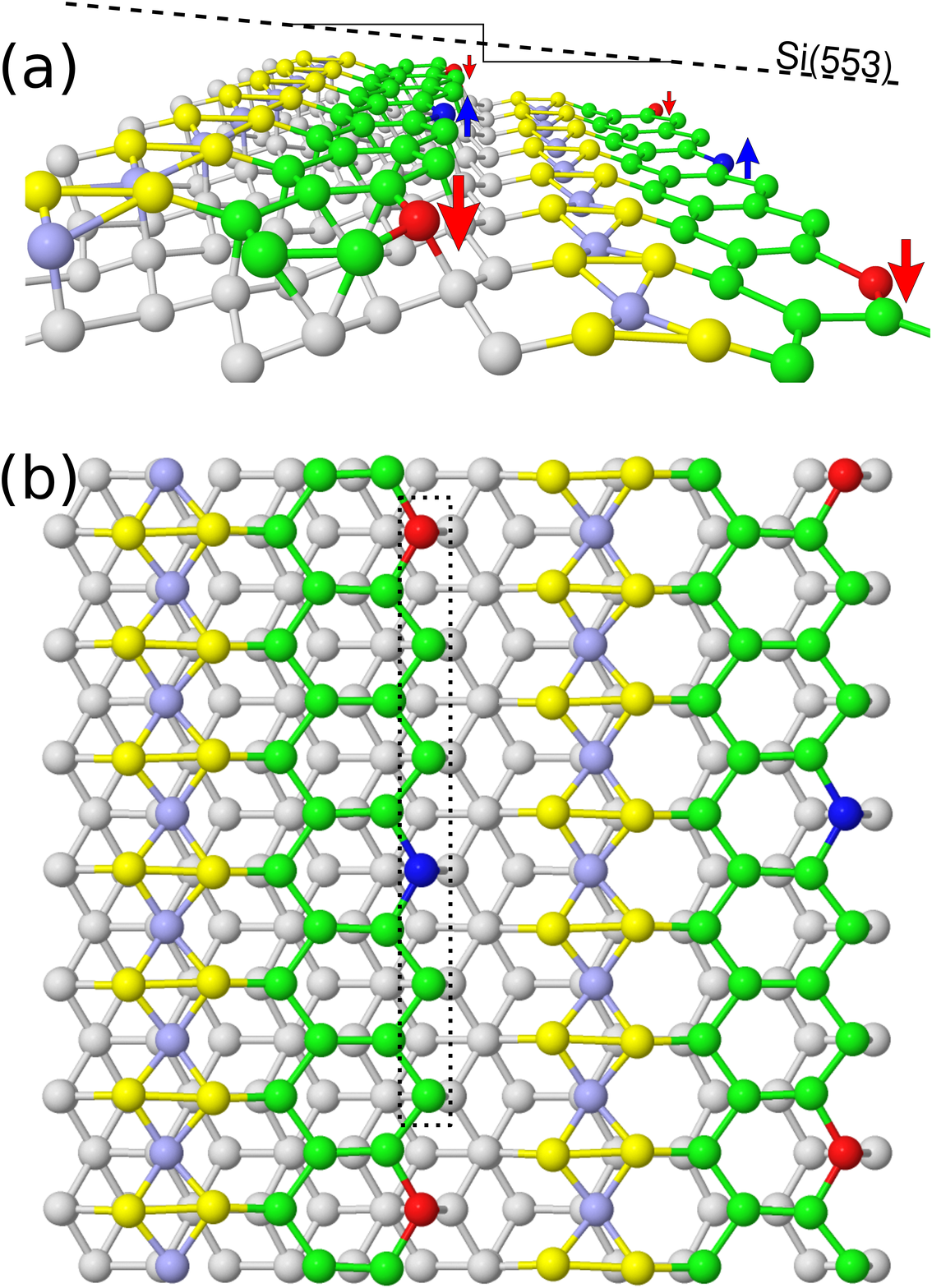}
\caption{(Color online) (a,b) Perspective and top views of Si(553)-Au
  in its electronic ground state. Yellow atoms are
Au, all others are Si. The Au atoms are embedded in flat terraces, which
are separated by steps consisting of 
Si atoms arranged as a honeycomb chain (green).
Every third Si atom (red, blue) at the step has a spin
magnetic moment of one Bohr magneton ($S=1/2$, arrows) from the
complete polarization of the electron occupying the dangling-bond 
orbital. The sign of the polarization (red vs.\ blue) alternates along the step.
The six atoms in the outlined box are the focus of the
ab-initio molecular dynamics discussed in Sec.\ III.
\label{model}}
\end{figure}

The physical and magnetic structure of Si(553)-Au in its ground state
were first proposed and discussed in Ref.\ \onlinecite{Erwin2010} and
for reference are reproduced in Fig.\ \ref{model}. This is a stepped surface
consisting of (111) terraces and bilayer steps, and is stabilized by
Au atoms that substitute for Si atoms in the surface layer of the
terrace.  The steps themselves consist of Si atoms organized into a
thin graphitic strip of honeycomb hexagons (the green atoms in Fig.\
\ref{model}).

% FIG1

The surface electronic structure of Si(553)-Au has two main contributions. The
first consists of two intense quasi-1D parabolic electron bands
centered at the boundary of the surface Brillouin zone. These ``Au
bands'' arise from the bonding and antibonding combinations of Au 6$s$
and subsurface Si orbitals (purple atoms). The bonding Au band is
approximately half-filled and the antibonding band approximately
one-fourth filled. 

The second contribution arises from the very edge of the Si honeycomb
chain, which consists of threefold-coordinated Si atoms. The
unpassivated $sp^3$ orbitals of these atoms can in principle be
occupied by zero, one, or two electrons.  The Si atoms themselves
supply, on average, one electron per orbital. The step edge does not
necessarily maintain this average occupancy, because electronic charge
can also be transferred to or from the Au bands. Indeed, DFT
calculations predict that the lowest energy configuration has one
electron in every third orbital (red and blue atoms in Fig.\
\ref{model}) and double occupancy everywhere else (green atoms). The
singly occupied orbitals are completely spin-polarized and hence have
local spin moments of 1 bohr magneton each. Physically, these atoms
relax slightly downward, by 0.3 \AA, compared to
their nonpolarized neighbors.  The sign of the spins alternates along
the step edge, with antiferromagnetic order favored by 15 meV per spin
relative to ferromagnetic order. Therefore the magnetic periodicity is
6$a_0$, where $a_0$ is the Si surface lattice constant. This is also
the smallest period that allows for the coexistence of 3$a_0$ spacing
of the spins and 2$a_0$ spacing (period doubling) within the Au chain.
This coexistence was first observed in STM
experiments\cite{Ahn2005,Snijders2006} and emerges naturally in DFT
calculations---but only when the spin degree of freedom is
unconstrained.\cite{Erwin2010}

\section{Finite temperature dynamics}\label{finitetemp}

The remainder of this article explores excitations of Si(553)-Au from
its ground state due to finite temperature.  Two main theoretical
tools were used: ab-initio molecular dynamics (MD) and kinetic Monte
Carlo (kMC) simulations. The first was used to identify the most
important low-energy activated processes and to determine their
activation barriers.  Because of the complexity of the system only
small time scales (tens of ps) and a small (1$\times$6) simulation
cell could be addressed using ab-initio MD. To reach much longer time
scales (tens of ns) and larger system sizes (up to 128 spins) we
constructed a one-dimensional kMC model based on the processes and
barriers determined from ab-initio MD.  In particular, the kMC model
allowed us to investigate finite-temperature behavior in the
presence of pinning defects---providing useful insight into
temperature-dependent results from scanning probe experiments, where
defects often play a critical role.

Two simplifying assumptions were used throughout this work. (1)
Electronic excitations were not considered, and consequently the
system stays on the Born-Oppenheimer surface. This assumption is
reasonable in view of the modest temperatures---room temperature and
lower---considered here.  (2) Spin flips were not allowed. Although,
as we will see below, the spins can diffuse among the Si step-edge
atoms, their signs and ordering remained that of the original
antiferromagnetic ordering.  Although a different initial spin
ordering might affect some details of the simulation, the overall
qualitative findings would be very similar.

\subsection{Ab-initio molecular dynamics}\label{md}

The MD simulations were performed using the same basic geometry and
computational parameters described in Ref.\ \onlinecite{Erwin2010}.
The Si(553)-Au surface was represented by six layers of Si plus the
reconstructed top surface layer and a vacuum region of 10 \AA.  All
atoms were free to move during the simulation except the bottom Si
layer and its passivating hydrogen layer. Total energies and forces
were calculated within the generalized-gradient approximation of
Perdew, Burke, and Ernzerhof to DFT using projector-augmented wave
potentials, as implemented in
VASP.\cite{Kresse1993,Kresse1996,Blochl1994,Kresse1999} The plane-wave
cutoff was 200 eV and only the $\Gamma$ point was used. The dynamics
simulations were performed in the canonical ensemble using a Nos\'{e}
thermostat and a time step of 3 fs. Five temperatures, equally spaced
in $1/T$, were used (57, 67, 80, 110, 133 K). For each temperature a
thermalization run of 10 ps was first performed, followed by a
dynamics run of 20 ps.

\begin{figure}[tb]
\includegraphics[width=8cm]{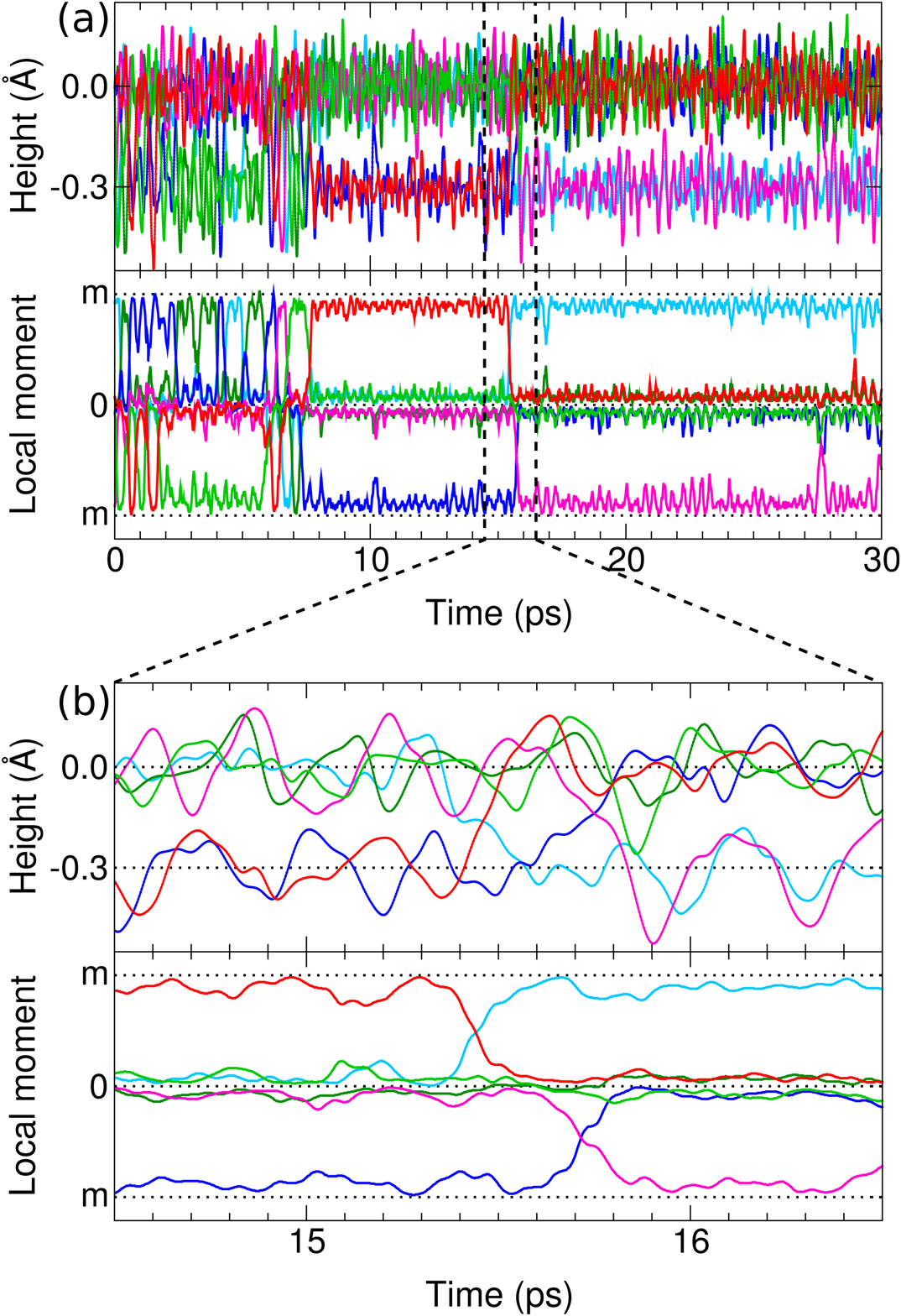}
\caption{(Color online) (a) Ab-initio molecular dynamics trajectories
  of the six Si
  step-edge atoms outlined in Fig.\ 1, at 57 K. Red and blue curves denote
  the red and blue atoms, which are initially spin-polarized. Other colors (magenta, cyan, dark green, light
  green) denote initially non-polarized atoms. Thermalization
  is completed by about 10 ps. Upper panel: height of each atom,
  relative to the average height of nonpolarized atoms. Lower panel: local spin moment of each
  atom. (b) Expanded view of two spin hops occurring at 15.44 ps (from the red atom to
  the cyan atom) and at 15.71 ps (from the blue atom to the magenta atom).
  \label{stepheightmom}}
\end{figure}

Figure \ref{stepheightmom} shows the resulting atomic trajectories
during the entire run of 10$^4$ MD time steps for the lowest
temperature studied, 57 K. The six curves are for the six Si step-edge
atoms in the outlined box of Fig.\ \ref{model}(b). The upper and lower
panels show the relative heights of the atoms and their local spin
moments, respectively.  After thermalization was achieved 
the system settled into its ground state configuration with two
spin-polarized atoms (red and blue) sitting $\sim$0.3 \AA\ lower than
their four non-polarized neighbors. 

% FIG2

The expanded view in Fig.\ \ref{stepheightmom}(b) focuses on two
events that occurred between 15 and 16 ps.  At 15.44 ps the magnitude
of the moment on the spin-up red atom went rapidly to zero while,
concurrently, a spin-up moment rapidly developed on the neighboring
cyan atom. At essentially the same time the height of the red atom
increased by 0.3 \AA\ to that of a non-polarized atom, while the cyan
atom moved down by the same amount. In summary, the spin-up moment that was
localized on the red atom hopped to one of its neighbors.

Very soon after, a second hop occurred at 15.71 ps. This
hop was made by the other spin (with the opposite sign) which moved
from the blue atom to the magenta atom. It is not a coincidence that
this hop occurred so soon after the first. The first hop changed the
minimum spacing between spins from 3$a_0$ to 2$a_0$, incurring an
energy penalty (discussed in detail below). This increase in energy in
turn reduced the barrier for any hop that restores the spacing to its
optimal value. The cyan and magenta atoms are indeed separated by
3$a_0$, and thus after two rapid spin hops the system was restored
to an equivalent ground state configuration, in which it remained for the
rest of the simulation.

The very small number of hops observed at 57 K makes it clear that
ab-initio MD simulations of Si(553)-Au at still lower temperatures, where
many of the relevant experiments are conducted, are not feasible.
Instead we turn to higher temperatures and ask how the frequency of
hopping events depends on temperature. This information will be useful
in Sec.\ \ref{kmc} for calibrating and validating our kMC model in a
temperature range accessible to both methods.

\begin{figure}[tb]
\includegraphics[width=8cm]{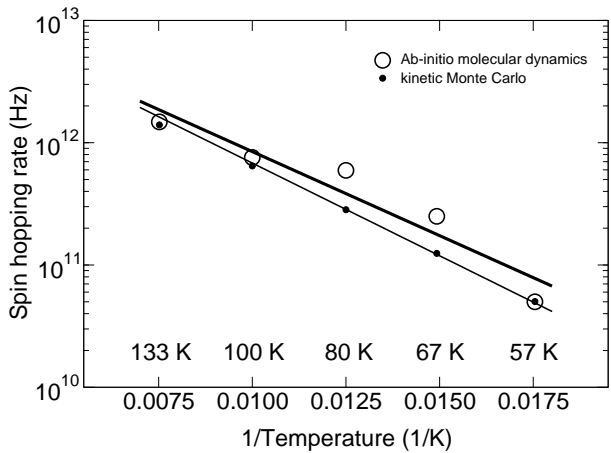}
\caption{
Temperature dependence of the hopping rate for Si spins along the
Si(553)-Au step edge.  The rates are time averages extracted from
ab-initio molecular dynamics (MD) trajectories, and are compared to
rates from kinetic Monte Carlo (kMC) simulations. The linear fits
describe Arrhenius behavior. The fit to ab-initio MD rates (thick
line) gives a pre-exponential factor $2.0\times 10^{13}$ s$^{-1}$ and
activation barrier 12 meV. For kMC rates (thin line) the values are
$2.2\times 10^{13}$ s$^{-1}$ and 13 meV.
\label{rates}}
\end{figure}

Figure \ref{rates} shows the resulting time-averaged hopping rate for a
single spin, versus inverse temperature. The rates for low
temperatures have large statistical uncertainties (not shown) and
hence it is reasonable to describe these results by a simple linear Arrhenius
fit, as shown. The attempt frequency, $2.0\times 10^{13}$ s$^{-1}$, is
on the order of a surface vibrational frequency, as expected. The
activation energy, 12 meV, represents a characteristic average of the
individual barriers for spin hops weighted by their relative probability
of occurrence.

% FIG3

\subsection{Kinetic Monte Carlo model}\label{kmc}

To construct the kMC model one first needs to enumerate all the
relevant spin hops and their individual rates. We used DFT
results obtained from the full Si(553)-Au system for this task. In the spirit
of simplicity we constructed the kMC model itself to be strictly
one-dimensional, with an arbitrarily large unit cell and periodic
boundary conditions. Thus the kMC simulations inherit much of the
accuracy of the DFT calculations but make the additional approximation
that spin hops along different step edges are independent.

\begin{figure}[tb]
\includegraphics[width=8cm]{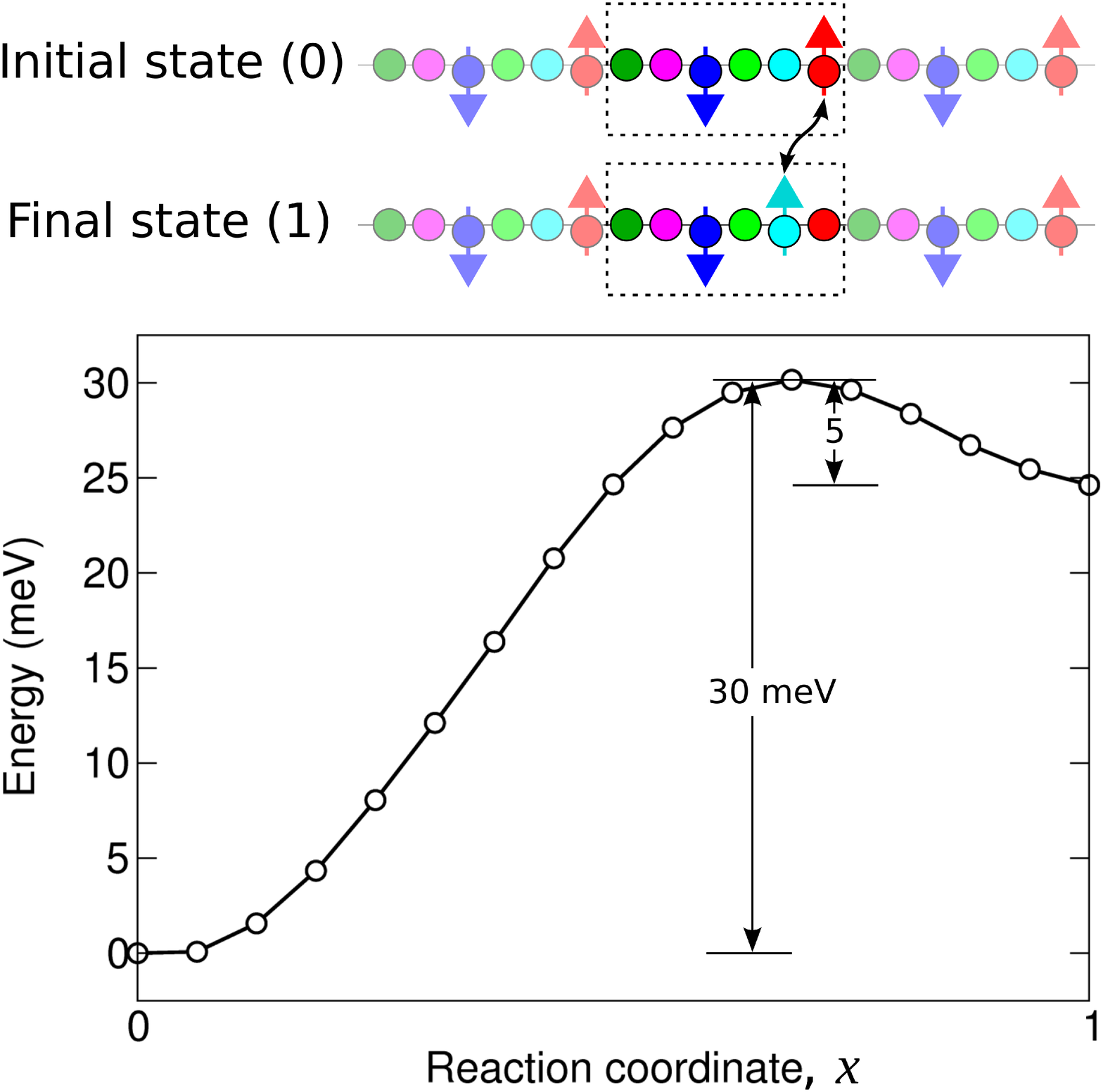}
\caption{
DFT potential energy surface for a single spin hopping from the red atom
to the neighboring cyan atom. 
The initial state (0) is the ground state depicted in Fig. 1.
The final state (1) is the metastable state, fully relaxed, that
exists between 15.44 and 15.71 ps in the MD simulation of Fig. 2. The
activation barrier is 30 meV for the forward reaction and 5 meV for
the reverse reaction. Atom colors correspond to the trajectories in
Fig. 2.
\label{barriers}}
\end{figure}

Figure \ref{barriers} shows the DFT potential energy surface for the
spin hop observed in Fig. \ref{stepheightmom}(b) at 15.44 ps into the
MD simulation.  Because the spins and the heights of the atoms are
tightly linked, the reaction coordinate $x$ is approximately given by
the relative heights $h$ of the red and cyan atoms, 
\begin{equation}
\label{eq:x}
x\approx[1-(h_{\rm cyan}-h_{\rm red})/\Delta h]/2,
\end{equation}
where $\Delta h= 0.3$ \AA\
is the equilibrium height difference between spin-polarized and
non-polarized atoms. To definitively determine the detailed reaction
pathway and potential energy surface we used the nudged elastic-band method.

% FIG4

This potential energy surface confirms the assertion, made in Sec.
\ref{md}, that the red-to-cyan (forward) spin hop incurs an energy
penalty that leads to a smaller barrier for the
cyan-to-red (reverse) hop. Specifically, the activation barrier for the
forward hop is 30 meV, the resulting energy penalty is 25
meV, and the barrier for the reverse hop is 5
meV. These two types of hops, and their calculated barriers, are two
of the three fundamental processes included in our kMC model.

For convenience we define here a more compact notation for enumerating
the different types of spin hops. Careful examination of the ab-initio
MD trajectories reveals that all spin hops were to an adjacent site;
there were no double hops. Hence we can label every hop as either
leftward ($\leftarrow$) or rightward ($\rightarrow$). We assume that
the barrier for a spin hop depends only on the spin's immediate
environment, that is, on the distances $ma_0$ and $na_0$ to the left
and right neighboring spins, respectively, measured before making the
hop. Using this notation we can express the barriers for the two hops
shown in Fig.\ \ref{barriers} as $b(3,3,\leftarrow)= 30$ meV and
$b(2,4,\rightarrow) = 5$ meV, where the two numerical arguments denote $m$
and $n$, respectively.

The third important hop we considered occurs when $m+n=5$, rather than
6 as depicted in Fig. \ref{barriers}. From DFT nudged elastic-band
calculations we find $b(3,2,\leftarrow)$ = $b(2,3,\rightarrow)= 14$ meV
(the barriers are equal by symmetry).  As expected from the
distances to the neighboring spins, this barrier is in between the
previous two.

The MD trajectories also show that two spins never occupy adjacent
sites. Because of this, our enumeration of the possible spin hops is
already complete for all cases with $m+n\le 6$. (It is worth noting
that the configuration in which a spin has both neighbors at $2a_0$ is
allowed, but because its adjacent sites cannot be occupied this spin
cannot hop until one of its neighbors does.)  The cases with $m+n\ge7$ are difficult to treat within
DFT but occur more rarely and thus are less important. For this reason
we treated the effect of neighbors beyond $3a_0$ as negligible, used
the barrier of 30 meV for any hop that brings a spin within $2a_0$ of
its neighbor, and assigned a single (arbitrary) barrier of 10 meV to
hops that maintain larger separations than this. This completes our
enumeration.

To finish the construction of the kMC model, we assumed that the rates
for all allowed spin hops are given by $r=a\exp(-b/kT)$, where $a$ is
a common prefactor and $b=b(m,n,\leftarrow)$ and
$b(m,n,\rightarrow)$ are the DFT barriers.

To determine the optimal value of $a$ and compare the predictions of
the kMC model to the ab-initio MD results, we applied the model to
the system discussed in Sec.\ \ref{md}---two
spin-polarized atoms in a six-atom unit cell with periodic boundary
conditions. The resulting kMC spin hopping rates
obtained using $a=6\times 10^{12}$ s$^{-1}$ are plotted in
Fig.\ \ref{rates} for direct comparison with the rates from ab-initio
MD. The kMC rates have negligible statistical errors and it is clear
that a simple Arrhenius fit describes them very well. Moreover, the 
fitted attempt frequency, $2.2\times 10^{13}$ s$^{-1}$, and
activation energy, 12 meV, are within 10\% of the MD
values. This confirms that the kMC model accurately reproduces the
ab-initio results within the temperature range considered.

\subsection{Spins at finite temperature near a defect}
\label{withdefects}

In real systems, the behavior of collective phenomena is often
controlled by defects that pin the phase of a low-temperature state
having broken symmetry.  On the Si(553)-Au surface, a variety of
defects---missing atoms, absorbates, etc.---have been observed to act
as pinning sites that locally stabilize the 1$\times$3 ground
state.\cite{Snijders2006,Kang2009b,Hasegawa2010,Shin2012} The
important role played by such pinning defects motivates our first
application of the kMC model.

We prepared a system consisting of 128 independent spins, with
periodic boundary conditions, initially arranged in the
antiferromagnetic ground state with uniform 3$a_0$ spacing. One of the
spins (at position 0) was pinned in place throughout the
simulation, thus representing a generic immobile defect. The
spins were allowed to hop stochastically among the 3$\times$128
lattice sites according to probabilities defined by the
hopping rates $r$.

\begin{figure}[tb]
\includegraphics[width=8cm]{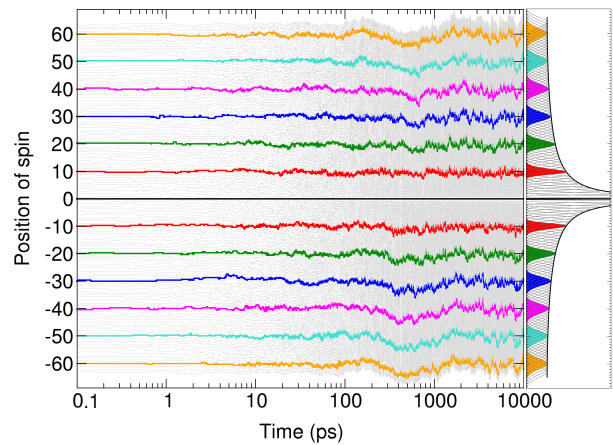}
\caption{(Color online) Kinetic Monte Carlo trajectories of 128 spins
  at 300 K in the presence of a pinning defect at the origin. Every
  tenth trace is colored for clarity. Right panel: histogram of the
  positions occupied  by each spin, weighted by the time spent there,
  obtained over a simulation time of 1 $\mu$s. The heavy curve is the
  envelope function $d^{-2/3}$ describing the decay of histogram heights with distance $d$
  from the pinning defect.
  \label{trajdefect}}
\end{figure}

Figure \ref{trajdefect} displays the resulting trajectories at 300 K
of all the spins over the first 10 ns of the simulation. For clarity
every tenth trajectory trace is colored.
As the system evolved, each spin explored a region of the
lattice around its initial position. These explorations were
relatively small for spins near the pinning defect and
became progressively larger for spins farther away.

% FIG5

The right panel in Fig.\ \ref{trajdefect} examines this thermally
induced wandering in greater detail. For each of the 128 spins a
histogram was made representing the position of that
spin at 300 K. At this temperature a simulation time of 1 $\mu$s was
sufficient to obtain the steady-state distribution. It is readily
apparent from examining the colored histograms that each is well
described by a gaussian function centered on the spin's initial
position.  Thus each of these gaussians is entirely specified by its
variance $\sigma^2$, whose value depends on the distance $d$ to the
pinning defect. To deduce this dependence we first note that the area
under each gaussian is by construction the same.  Hence the height of each gaussian is
proportional to $1/\sigma$.  We find empirically that the dependence
of these heights on distance is given with excellent accuracy as
$d^{-2/3}$. An envelope function with this dependence is shown on the
histogram plot as a heavy black curve. From this dependence we thus
deduce that the thermally induced widths $w$, defined here as
$2\sigma$, increase with distance from a pinning defect as
$w\sim d^{2/3}$.

Now we move on to explore how temperature affects the thermal
wandering of spins near a pinning defect. We repeated the kMC simulation
and analysis in Fig.\ \ref{trajdefect} for a series of temperatures
between 10 and 300 K. We focus on the variation of the thermal widths
$w$ as a function of temperature $T$. 

% FIG6
\begin{figure}[tb]
\includegraphics[width=8cm]{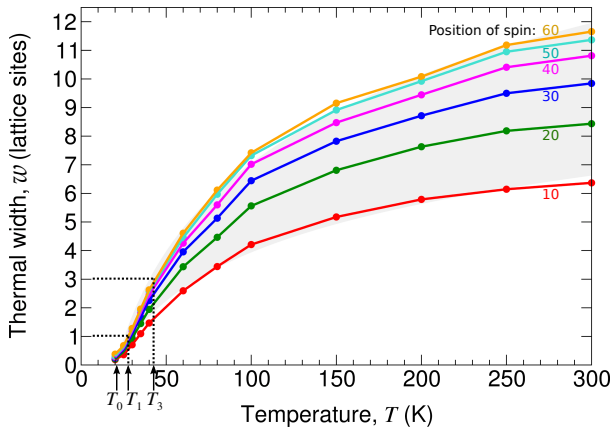}
\caption{(Color online) 
Temperature dependence of the thermal widths $w(T)$ of every tenth
spin, in the presence of a pinning defect. Labels indicate the spin's
distance from the defect, in units of 3$a_0$. Colors correspond to the
trajectories in Fig. 5. The light gray shaded area is bounded by
logarithmic fits to the lower (red) and upper (orange) data points.
The characteristic temperatures $T_0$, $T_1$, and $T_3$ describe
different criteria by which thermal wandering of spins is expected to
be either eliminated or suppressed; see discussion in text.
\label{widthsdefect}}
\end{figure}

Figure \ref{widthsdefect} summarizes the resulting temperature
dependence. The six datasets show $w(T)$ for every tenth spin of the
128-spin simulation cell. For reference, the six values at $T=300$ K
correspond to the six gaussian widths in the upper half of the
histogram panel in Fig.\ \ref{trajdefect}. We find empirically that
the dependence of each dataset on temperature is close to logarithmic,
as shown by the light gray shaded area. This implies that
the dependence of the thermal widths on distance and temperature
can be separated and written as 
\begin{equation}
\label{eq:w}
w(d,T)=w_0(d)\ln(T/T_0),   
\end{equation}
where $w_0(d)\sim d^{2/3}$ and the characteristic temperature $T_0$
has the fitted value 21 K. All thermal wandering is, by definition,
completely eliminated at $T_0$.  But two less restrictive criteria may
be more relevant for interpreting the experimentally observed
transition to the period-tripled ground state. At $T_1=27$ K the
thermal widths for all 128 spins become smaller than the width of a
single lattice site. Hence, below this temperature the spins will in
effect be frozen into place on every third lattice site.  At still
higher temperature, $T_3=42$ K, all thermal widths are less than or
equal to the average spacing (three lattice sites) between
spins. Hence the spins will first become distinguishable as the system
is cooled below this temperature.

To generalize this result and make predictions that can be tested by
experiment, we first assume that real systems can be characterized by
a known average concentration $c$ of pinning defects. Because the
defects are distributed in 1D, the characteristic distance $d$ from a
spin to the nearest defect scales as $1/c$. By inverting Eq.\
\ref{eq:w} we then immediately obtain a simple result: the
temperatures $T_{1,3}$ at which all spins in the system become either
frozen into place or distinguishable will scale as $T_{1,3} \sim
\exp(c^{2/3})$. Thus we predict this scaling to describe the
temperature at which the period-tripled ground state of Si(553)-Au is
first observed.

By inserting into this qualitative relationship the appropriate
constants obtained from the kMC simulations, we derive a
quantitative prediction for the freezing temperature,
\begin{equation}
  \label{eq:T}
  T_1=T_0 \exp(k\,c^{2/3}),
\end{equation}
where $k=11.8$ is a dimensionless constant and $c$ is expressed in the
dimensionless units of defects per lattice site. The corresponding
equation for $T_3$ can be obtained from Eq.\ \ref{eq:T} by multiplying
the argument of the exponential by three. A useful guide to
understanding the importance of defects in Si(553)-Au is provided by
linearizing Eq.\ \ref{eq:T} around a physically plausible value
(10$^{-2}$) for the defect concentration.  This leads to the result
that a change in the defect concentration will raise
the freezing temperature by $\Delta T_1 = \gamma\Delta c$, with
proportionality constant $\gamma=1330$ K. Thus, for samples with
approximately one
defect every 100 lattice sites, a doubling of this 
concentration will increase the freezing
temperature by 13 K.

\subsection{Spins at finite temperature in the absence of defects}
\label{withoutdefects}

Although a system completely free of defects is obviously unrealistic,
the behavior of such an idealized system nevertheless offers
complementary insight into the thermal wandering of spins when the
concentration of defects is very low. As we show below, despite the
absence of defects, the statistical behavior of the spins still
exhibits a sudden and qualitative change at about 30 K.

% FIG7
\begin{figure}[tb]
\includegraphics[width=8cm]{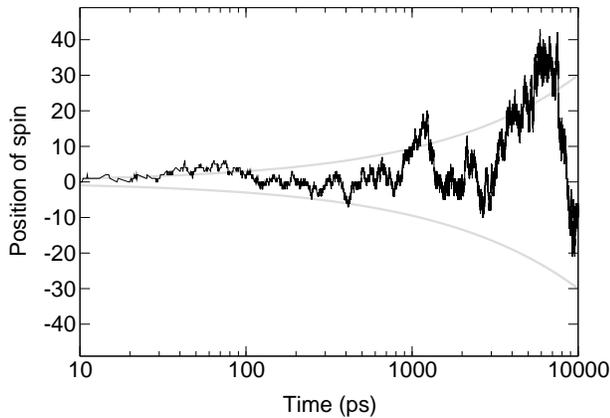}
\caption{Trajectory of a single spin at 100 K in the absence of
  pinning defects. Gray curve shows the theoretical average displacement
  versus time for a isolated random walker in one dimension, $\langle d\rangle \propto \sqrt{t}$.
  \label{randomwalk}}
\end{figure}

We constructed a periodic system of 64 spins similar to that described in
Sec.\ \ref{withdefects}, but now without a pinning defect. Thus each
spin executed a random walk in 1D. Figure \ref{randomwalk} shows
a typical trajectory at 100 K for one of the 64 spins. Despite the
stochastic nature of this single trajectory, it is already plausible
that the average displacement $\langle d\rangle$ depends on the time
$t$ according to $\langle d\rangle \propto \sqrt{t}$, which is the well-known
result for a single unbiased random walker in one dimension.

% FIG8
\begin{figure}[b]
\includegraphics[width=8cm]{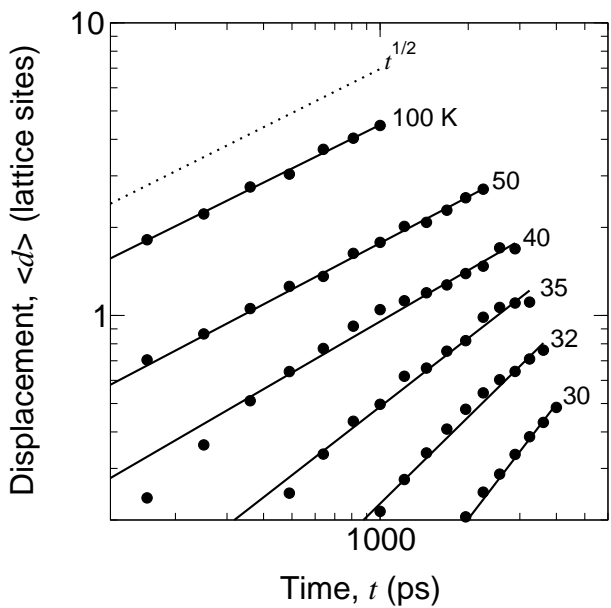}
\caption{
Average displacement versus time of a single spin in the absence of
defects, at the indicated temperatures. Circles are 
statistical averages over 2000 kMC simulations.
Straight lines are fits to $t^H$, where $H$ is
the Hurst exponent. The dotted line indicates $H$=1/2.
\label{avgdisplacement}}
\end{figure}

To analyze this behavior more systematically, we performed many
independent kMC simulations and computed the average displacements as a function of
time. Figure \ref{avgdisplacement} shows these averages on a log-log
scale for a range of temperatures. At high temperatures we indeed
obtain the behavior $\langle d\rangle \propto t^{1/2}$, which is indicated by
the dotted line. This behavior persists until the temperature reaches the
range 30--35 K, where it still exhibits power-law behavior
$\langle d\rangle \propto t^H$ but with a progressively larger
exponent $H>1/2$.

% FIG9
\begin{figure}[tb]
\includegraphics[width=8cm]{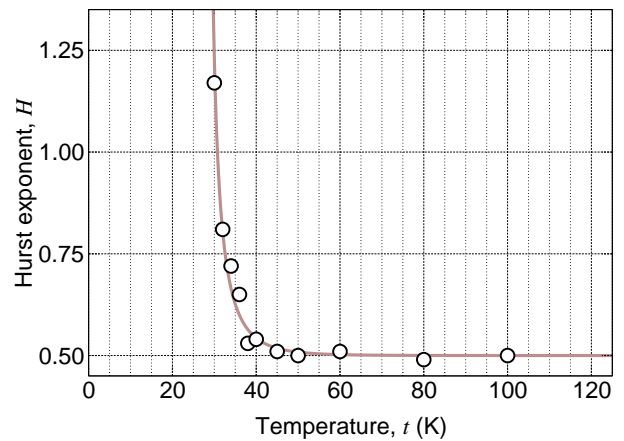}
\caption{
Hurst exponent versus temperature for spins in the absence of a defect. The solid curve is a fit to a
generalized susceptibility with a characteristic temperature $T_c=28$ K.
\label{hurst}}
\end{figure}

In general, a system for which the long-time displacements are
characterized by a Hurst exponent $H=1/2$ is said to be uncorrelated,
while $H>1/2$ indicates that long-time correlations are
present.\cite{Rangarajan2000} Figure \ref{hurst} shows the Hurst
exponent $H$, obtained by fitting the time-averaged displacements, as
a function of temperature. Above $\sim$40 K the system displays
uncorrelated behavior. Below this temperature we observe the rapid
onset of correlated behavior.  To quantify the temperature of this
transition we fit the Hurst exponents to a generalized susceptibility
of the form $H=(1/2)/[1-(T_c/T)^\nu]$ and obtain a characteristic
temperature $T_c=28$ K. Hence as the clean system is cooled toward
$T_c$ the random walks described by individual spins rapidly lose
their independent character. This transition occurs with a
characteristic temperature comparable to that obtained, $T_0=21$ K, by
extrapolating from the behavior in the presence of defects. Thus these
two complementary approaches lead to qualitatively as well as
quantitatively similar conclusions.

\section{Discussion and conclusions}

A guiding principle in one-dimensional physics is provided by the
Mermin-Wagner theorem, which states that phase transitions cannot
occur above $T=0$ if the interactions are
short-ranged.\cite{Mermin1966} Exploring the different manifestations
of this theorem in real systems can yield new and unanticipated
insights.  For example, a previous publication by one of us
demonstrated theoretically that when the interactions are $not$
short-ranged, as for $1/r$ Coulomb interactions, then a well-defined
thermodynamic phase transition can indeed occur---and likely does
occur for a system of Ba adsorbates on
Si(111).\cite{Erwin2005a,Bruch2005} In the present article the new
insights into 1D physics are of a different kind: we have shown that
even when the interactions are short-ranged, and the system purely 1D,
a well-ordered phase with the clear signature of a broken symmetry can
form well above $T=0$. Moreover, an approximate transition temperature
can be readily identified using realistic simulations and straightforward
statistical analysis.

For Si(553)-Au the precise nature of the interacting entities is
unexpected and somewhat subtle: our MD simulations showed that they are
neither simple vibrations of atoms, nor spins on a simple fixed
lattice, but rather a tightly coupled combination of both. In this
sense a description of Si(553)-Au based on spin polarons is
appropriate. Our kMC simulations showed that as the temperature of the
system is raised, the ground state $3a_0$ crystal formed by these
polarons melts at a temperature we estimate to be $\sim$30 K for
perfectly clean systems, and higher for systems with pinning defects.
A direct experimental test of this description is afforded by Eq.\
\ref{eq:T}, which predicts how the transition temperature varies with
the concentration of defects.

It is important to acknowledge some limitations of this work. Because our
focus has been on the behavior of Si(553)-Au at low temperature, we
have assumed that the spin-polarized silicon states remain polarized
at higher temperatures. Our preliminary calculations indicate that
this assumption is completely justified at the temperatures of
interest here. However, at much higher temperatures the thermally
induced vibrations of the step edge atoms increasingly render the
system nonpolarized for part of the time. For example, at room
temperature the average spin moment is reduced to roughly 2/3 of its
low-temperature value of 1 bohr magneton. Future theoretical
investigations into the behavior of Si(553)-Au near room temperature
will have to account for this thermal suppression of the spin
polarization.

Finally, as mentioned in Sec.\ \ref{finitetemp}, we have throughout
assumed for simplicity that the ordering of the spins remains
antiferromagnetic, as in the ground state. Preliminary calculations
show that the barriers for spin hopping depend quantitatively,
although not qualitatively, on the signs of the neighboring spins. A
generalization of our kMC model that includes spin flips would be a
very interesting direction to pursue, but we anticipate that the
qualitative conclusions drawn here---as well as the overall
consistency between our findings and those of existing
experiments---would be largely unchanged.

\begin{acknowledgments}
  Many discussions with F.J. Himpsel are gratefully acknowledged.
  This work was supported by the Office of Naval Research (SCE) and the
  Department of Energy, Basic Energy Sciences, Materials Sciences and
  Engineering Division (PCS). Computations were performed at the DoD
  Major Shared Resource Centers at AFRL and ERDC.
\end{acknowledgments}

%\bibliography{../../databaseNew}

\begin{thebibliography}{35}%
\makeatletter
\providecommand \@ifxundefined [1]{%
 \@ifx{#1\undefined}
}%
\providecommand \@ifnum [1]{%
 \ifnum #1\expandafter \@firstoftwo
 \else \expandafter \@secondoftwo
 \fi
}%
\providecommand \@ifx [1]{%
 \ifx #1\expandafter \@firstoftwo
 \else \expandafter \@secondoftwo
 \fi
}%
\providecommand \natexlab [1]{#1}%
\providecommand \enquote  [1]{``#1''}%
\providecommand \bibnamefont  [1]{#1}%
\providecommand \bibfnamefont [1]{#1}%
\providecommand \citenamefont [1]{#1}%
\providecommand \href@noop [0]{\@secondoftwo}%
\providecommand \href [0]{\begingroup \@sanitize@url \@href}%
\providecommand \@href[1]{\@@startlink{#1}\@@href}%
\providecommand \@@href[1]{\endgroup#1\@@endlink}%
\providecommand \@sanitize@url [0]{\catcode `\\12\catcode `\$12\catcode
  `\&12\catcode `\#12\catcode `\^12\catcode `\_12\catcode `\%12\relax}%
\providecommand \@@startlink[1]{}%
\providecommand \@@endlink[0]{}%
\providecommand \url  [0]{\begingroup\@sanitize@url \@url }%
\providecommand \@url [1]{\endgroup\@href {#1}{\urlprefix }}%
\providecommand \urlprefix  [0]{URL }%
\providecommand \Eprint [0]{\href }%
\providecommand \doibase [0]{http://dx.doi.org/}%
\providecommand \selectlanguage [0]{\@gobble}%
\providecommand \bibinfo  [0]{\@secondoftwo}%
\providecommand \bibfield  [0]{\@secondoftwo}%
\providecommand \translation [1]{[#1]}%
\providecommand \BibitemOpen [0]{}%
\providecommand \bibitemStop [0]{}%
\providecommand \bibitemNoStop [0]{.\EOS\space}%
\providecommand \EOS [0]{\spacefactor3000\relax}%
\providecommand \BibitemShut  [1]{\csname bibitem#1\endcsname}%
\let\auto@bib@innerbib\@empty
%</preamble>
\bibitem [{\citenamefont {Crain}\ \emph {et~al.}(2003)\citenamefont {Crain},
  \citenamefont {Kirakosian}, \citenamefont {Altmann}, \citenamefont
  {Bromberger}, \citenamefont {Erwin}, \citenamefont {McChesney}, \citenamefont
  {Lin},\ and\ \citenamefont {Himpsel}}]{Crain2003}%
  \BibitemOpen
  \bibfield  {author} {\bibinfo {author} {\bibfnamefont {J.~N.}\ \bibnamefont
  {Crain}}, \bibinfo {author} {\bibfnamefont {A.}~\bibnamefont {Kirakosian}},
  \bibinfo {author} {\bibfnamefont {K.~N.}\ \bibnamefont {Altmann}}, \bibinfo
  {author} {\bibfnamefont {C.}~\bibnamefont {Bromberger}}, \bibinfo {author}
  {\bibfnamefont {S.~C.}\ \bibnamefont {Erwin}}, \bibinfo {author}
  {\bibfnamefont {J.}~\bibnamefont {McChesney}}, \bibinfo {author}
  {\bibfnamefont {J.-L.}\ \bibnamefont {Lin}}, \ and\ \bibinfo {author}
  {\bibfnamefont {F.~J.}\ \bibnamefont {Himpsel}},\ }\href@noop {} {\bibfield
  {journal} {\bibinfo  {journal} {Phys. Rev. Lett.}\ }\textbf {\bibinfo
  {volume} {90}},\ \bibinfo {pages} {176805} (\bibinfo {year}
  {2003})}\BibitemShut {NoStop}%
\bibitem [{\citenamefont {Ghose}\ \emph {et~al.}(2005)\citenamefont {Ghose},
  \citenamefont {Robinson}, \citenamefont {Bennett},\ and\ \citenamefont
  {Himpsel}}]{Ghose2005}%
  \BibitemOpen
  \bibfield  {author} {\bibinfo {author} {\bibfnamefont {S.~K.}\ \bibnamefont
  {Ghose}}, \bibinfo {author} {\bibfnamefont {I.~K.}\ \bibnamefont {Robinson}},
  \bibinfo {author} {\bibfnamefont {P.~A.}\ \bibnamefont {Bennett}}, \ and\
  \bibinfo {author} {\bibfnamefont {F.~J.}\ \bibnamefont {Himpsel}},\ }\href
  {\doibase 10.1016/j.susc.2005.02.053} {\bibfield  {journal} {\bibinfo
  {journal} {Surf. Sci.}\ }\textbf {\bibinfo {volume} {581}},\ \bibinfo {pages}
  {199} (\bibinfo {year} {2005})}\BibitemShut {NoStop}%
\bibitem [{\citenamefont {Takayama}\ \emph {et~al.}(2009)\citenamefont
  {Takayama}, \citenamefont {Voegeli}, \citenamefont {Shirasawa}, \citenamefont
  {Kubo}, \citenamefont {Abe}, \citenamefont {Takahashi}, \citenamefont
  {Akimoto},\ and\ \citenamefont {Sugiyama}}]{Takayama2009}%
  \BibitemOpen
  \bibfield  {author} {\bibinfo {author} {\bibfnamefont {T.}~\bibnamefont
  {Takayama}}, \bibinfo {author} {\bibfnamefont {W.}~\bibnamefont {Voegeli}},
  \bibinfo {author} {\bibfnamefont {T.}~\bibnamefont {Shirasawa}}, \bibinfo
  {author} {\bibfnamefont {K.}~\bibnamefont {Kubo}}, \bibinfo {author}
  {\bibfnamefont {M.}~\bibnamefont {Abe}}, \bibinfo {author} {\bibfnamefont
  {T.}~\bibnamefont {Takahashi}}, \bibinfo {author} {\bibfnamefont
  {K.}~\bibnamefont {Akimoto}}, \ and\ \bibinfo {author} {\bibfnamefont
  {H.}~\bibnamefont {Sugiyama}},\ }\href {\doibase 10.1380/ejssnt.2009.533}
  {\bibfield  {journal} {\bibinfo  {journal} {e-Journal of Surface Science and
  Nanotechnology}\ }\textbf {\bibinfo {volume} {7}},\ \bibinfo {pages} {533}
  (\bibinfo {year} {2009})}\BibitemShut {NoStop}%
\bibitem [{\citenamefont {Voegeli}\ \emph {et~al.}(2010)\citenamefont
  {Voegeli}, \citenamefont {Takayama}, \citenamefont {Shirasawa}, \citenamefont
  {Abe}, \citenamefont {Kubo}, \citenamefont {Takahashi}, \citenamefont
  {Akimoto},\ and\ \citenamefont {Sugiyama}}]{Voegeli2010}%
  \BibitemOpen
  \bibfield  {author} {\bibinfo {author} {\bibfnamefont {W.}~\bibnamefont
  {Voegeli}}, \bibinfo {author} {\bibfnamefont {T.}~\bibnamefont {Takayama}},
  \bibinfo {author} {\bibfnamefont {T.}~\bibnamefont {Shirasawa}}, \bibinfo
  {author} {\bibfnamefont {M.}~\bibnamefont {Abe}}, \bibinfo {author}
  {\bibfnamefont {K.}~\bibnamefont {Kubo}}, \bibinfo {author} {\bibfnamefont
  {T.}~\bibnamefont {Takahashi}}, \bibinfo {author} {\bibfnamefont
  {K.}~\bibnamefont {Akimoto}}, \ and\ \bibinfo {author} {\bibfnamefont
  {H.}~\bibnamefont {Sugiyama}},\ }\href {\doibase 10.1103/PhysRevB.82.075426}
  {\bibfield  {journal} {\bibinfo  {journal} {Physical Review B}\ }\textbf
  {\bibinfo {volume} {82}},\ \bibinfo {pages} {075426} (\bibinfo {year}
  {2010})}\BibitemShut {NoStop}%
\bibitem [{\citenamefont {Riikonen}\ and\ \citenamefont
  {Sanchez-Portal}(2005)}]{Riikonen2005}%
  \BibitemOpen
  \bibfield  {author} {\bibinfo {author} {\bibfnamefont {S.}~\bibnamefont
  {Riikonen}}\ and\ \bibinfo {author} {\bibfnamefont {D.}~\bibnamefont
  {Sanchez-Portal}},\ }\href {\doibase 10.1088/0957-4484/16/5/015} {\bibfield
  {journal} {\bibinfo  {journal} {Nanotechnology}\ }\textbf {\bibinfo {volume}
  {16}},\ \bibinfo {pages} {S218} (\bibinfo {year} {2005})}\BibitemShut
  {NoStop}%
\bibitem [{\citenamefont {Riikonen}\ and\ \citenamefont
  {Sanchez-Portal}(2006)}]{Riikonen2006}%
  \BibitemOpen
  \bibfield  {author} {\bibinfo {author} {\bibfnamefont {S.}~\bibnamefont
  {Riikonen}}\ and\ \bibinfo {author} {\bibfnamefont {D.}~\bibnamefont
  {Sanchez-Portal}},\ }\href {\doibase 10.1016/j.susc.2005.12.043} {\bibfield
  {journal} {\bibinfo  {journal} {Surf. Sci.}\ }\textbf {\bibinfo {volume}
  {600}},\ \bibinfo {pages} {1201} (\bibinfo {year} {2006})}\BibitemShut
  {NoStop}%
\bibitem [{\citenamefont {Riikonen}\ and\ \citenamefont
  {Sanchez-Portal}(2008)}]{Riikonen2008}%
  \BibitemOpen
  \bibfield  {author} {\bibinfo {author} {\bibfnamefont {S.}~\bibnamefont
  {Riikonen}}\ and\ \bibinfo {author} {\bibfnamefont {D.}~\bibnamefont
  {Sanchez-Portal}},\ }\href {\doibase 10.1103/PhysRevB.77.165418} {\bibfield
  {journal} {\bibinfo  {journal} {Physical Review B}\ }\textbf {\bibinfo
  {volume} {77}},\ \bibinfo {pages} {165418} (\bibinfo {year}
  {2008})}\BibitemShut {NoStop}%
\bibitem [{\citenamefont {Krawiec}(2010)}]{Krawiec2010}%
  \BibitemOpen
  \bibfield  {author} {\bibinfo {author} {\bibfnamefont {M.}~\bibnamefont
  {Krawiec}},\ }\href {\doibase 10.1103/PhysRevB.81.115436} {\bibfield
  {journal} {\bibinfo  {journal} {Physical Review B}\ }\textbf {\bibinfo
  {volume} {81}},\ \bibinfo {pages} {115436} (\bibinfo {year}
  {2010})}\BibitemShut {NoStop}%
\bibitem [{\citenamefont {Barke}\ \emph {et~al.}(2009)\citenamefont {Barke},
  \citenamefont {Zheng}, \citenamefont {Bockenhauer}, \citenamefont {Sell},
  \citenamefont {v.~Oeynhausen}, \citenamefont {Meiwes-Broer}, \citenamefont
  {Erwin},\ and\ \citenamefont {Himpsel}}]{Barke2009}%
  \BibitemOpen
  \bibfield  {author} {\bibinfo {author} {\bibfnamefont {I.}~\bibnamefont
  {Barke}}, \bibinfo {author} {\bibfnamefont {F.}~\bibnamefont {Zheng}},
  \bibinfo {author} {\bibfnamefont {S.}~\bibnamefont {Bockenhauer}}, \bibinfo
  {author} {\bibfnamefont {K.}~\bibnamefont {Sell}}, \bibinfo {author}
  {\bibfnamefont {V.}~\bibnamefont {v.~Oeynhausen}}, \bibinfo {author}
  {\bibfnamefont {K.~H.}\ \bibnamefont {Meiwes-Broer}}, \bibinfo {author}
  {\bibfnamefont {S.~C.}\ \bibnamefont {Erwin}}, \ and\ \bibinfo {author}
  {\bibfnamefont {F.~J.}\ \bibnamefont {Himpsel}},\ }\href {\doibase
  10.1103/PhysRevB.79.155301} {\bibfield  {journal} {\bibinfo  {journal}
  {Physical Review B}\ }\textbf {\bibinfo {volume} {79}},\ \bibinfo {pages}
  {155301} (\bibinfo {year} {2009})}\BibitemShut {NoStop}%
\bibitem [{\citenamefont {Crain}\ and\ \citenamefont
  {Pierce}(2005)}]{Crain2005}%
  \BibitemOpen
  \bibfield  {author} {\bibinfo {author} {\bibfnamefont {J.~N.}\ \bibnamefont
  {Crain}}\ and\ \bibinfo {author} {\bibfnamefont {D.~T.}\ \bibnamefont
  {Pierce}},\ }\href {\doibase 10.1126/science.1106911} {\bibfield  {journal}
  {\bibinfo  {journal} {Science}\ }\textbf {\bibinfo {volume} {307}},\ \bibinfo
  {pages} {703} (\bibinfo {year} {2005})}\BibitemShut {NoStop}%
\bibitem [{\citenamefont {Crain}\ \emph {et~al.}(2006)\citenamefont {Crain},
  \citenamefont {Stiles}, \citenamefont {Stroscio},\ and\ \citenamefont
  {Pierce}}]{Crain2006}%
  \BibitemOpen
  \bibfield  {author} {\bibinfo {author} {\bibfnamefont {J.~N.}\ \bibnamefont
  {Crain}}, \bibinfo {author} {\bibfnamefont {M.~D.}\ \bibnamefont {Stiles}},
  \bibinfo {author} {\bibfnamefont {J.~A.}\ \bibnamefont {Stroscio}}, \ and\
  \bibinfo {author} {\bibfnamefont {D.~T.}\ \bibnamefont {Pierce}},\ }\href
  {\doibase 10.1103/PhysRevLett.96.156801} {\bibfield  {journal} {\bibinfo
  {journal} {Phys. Rev. Lett.}\ }\textbf {\bibinfo {volume} {96}},\ \bibinfo
  {pages} {156801} (\bibinfo {year} {2006})}\BibitemShut {NoStop}%
\bibitem [{\citenamefont {Okino}\ \emph
  {et~al.}(2007{\natexlab{a}})\citenamefont {Okino}, \citenamefont {Matsuda},
  \citenamefont {Yamazaki}, \citenamefont {Hobara},\ and\ \citenamefont
  {Hasegawa}}]{Okino2007a}%
  \BibitemOpen
  \bibfield  {author} {\bibinfo {author} {\bibfnamefont {H.}~\bibnamefont
  {Okino}}, \bibinfo {author} {\bibfnamefont {I.}~\bibnamefont {Matsuda}},
  \bibinfo {author} {\bibfnamefont {S.}~\bibnamefont {Yamazaki}}, \bibinfo
  {author} {\bibfnamefont {R.}~\bibnamefont {Hobara}}, \ and\ \bibinfo {author}
  {\bibfnamefont {S.}~\bibnamefont {Hasegawa}},\ }\href {\doibase
  10.1103/PhysRevB.76.035424} {\bibfield  {journal} {\bibinfo  {journal}
  {Physical Review B}\ }\textbf {\bibinfo {volume} {76}},\ \bibinfo {pages}
  {035424} (\bibinfo {year} {2007}{\natexlab{a}})}\BibitemShut {NoStop}%
\bibitem [{\citenamefont {Ryang}\ \emph {et~al.}(2007)\citenamefont {Ryang},
  \citenamefont {Kang}, \citenamefont {Yeom},\ and\ \citenamefont
  {Jeong}}]{Ryang2007}%
  \BibitemOpen
  \bibfield  {author} {\bibinfo {author} {\bibfnamefont {K.-D.}\ \bibnamefont
  {Ryang}}, \bibinfo {author} {\bibfnamefont {P.~G.}\ \bibnamefont {Kang}},
  \bibinfo {author} {\bibfnamefont {H.~W.}\ \bibnamefont {Yeom}}, \ and\
  \bibinfo {author} {\bibfnamefont {S.}~\bibnamefont {Jeong}},\ }\href
  {\doibase 10.1103/PhysRevB.76.205325} {\bibfield  {journal} {\bibinfo
  {journal} {Physical Review B}\ }\textbf {\bibinfo {volume} {76}},\ \bibinfo
  {pages} {205325} (\bibinfo {year} {2007})}\BibitemShut {NoStop}%
\bibitem [{\citenamefont {Okino}\ \emph
  {et~al.}(2007{\natexlab{b}})\citenamefont {Okino}, \citenamefont {Matsuda},
  \citenamefont {Hobara}, \citenamefont {Hasegawa}, \citenamefont {Kim},\ and\
  \citenamefont {Lee}}]{Okino2007}%
  \BibitemOpen
  \bibfield  {author} {\bibinfo {author} {\bibfnamefont {H.}~\bibnamefont
  {Okino}}, \bibinfo {author} {\bibfnamefont {I.}~\bibnamefont {Matsuda}},
  \bibinfo {author} {\bibfnamefont {R.}~\bibnamefont {Hobara}}, \bibinfo
  {author} {\bibfnamefont {S.}~\bibnamefont {Hasegawa}}, \bibinfo {author}
  {\bibfnamefont {Y.}~\bibnamefont {Kim}}, \ and\ \bibinfo {author}
  {\bibfnamefont {G.}~\bibnamefont {Lee}},\ }\href {\doibase
  10.1103/PhysRevB.76.195418} {\bibfield  {journal} {\bibinfo  {journal}
  {Physical Review B}\ }\textbf {\bibinfo {volume} {76}},\ \bibinfo {pages}
  {195418} (\bibinfo {year} {2007}{\natexlab{b}})}\BibitemShut {NoStop}%
\bibitem [{\citenamefont {Ahn}\ \emph {et~al.}(2008)\citenamefont {Ahn},
  \citenamefont {Kang}, \citenamefont {Byun},\ and\ \citenamefont
  {Yeom}}]{Ahn2008}%
  \BibitemOpen
  \bibfield  {author} {\bibinfo {author} {\bibfnamefont {J.~R.}\ \bibnamefont
  {Ahn}}, \bibinfo {author} {\bibfnamefont {P.~G.}\ \bibnamefont {Kang}},
  \bibinfo {author} {\bibfnamefont {J.~H.}\ \bibnamefont {Byun}}, \ and\
  \bibinfo {author} {\bibfnamefont {H.~W.}\ \bibnamefont {Yeom}},\ }\href
  {\doibase 10.1103/PhysRevB.77.035401} {\bibfield  {journal} {\bibinfo
  {journal} {Physical Review B}\ }\textbf {\bibinfo {volume} {77}},\ \bibinfo
  {pages} {035401} (\bibinfo {year} {2008})}\BibitemShut {NoStop}%
\bibitem [{\citenamefont {Kang}\ \emph
  {et~al.}(2009{\natexlab{a}})\citenamefont {Kang}, \citenamefont {Jeong},\
  and\ \citenamefont {Yeom}}]{Kang2009}%
  \BibitemOpen
  \bibfield  {author} {\bibinfo {author} {\bibfnamefont {P.-G.}\ \bibnamefont
  {Kang}}, \bibinfo {author} {\bibfnamefont {H.}~\bibnamefont {Jeong}}, \ and\
  \bibinfo {author} {\bibfnamefont {H.~W.}\ \bibnamefont {Yeom}},\ }\href
  {\doibase 10.1103/PhysRevB.79.113403} {\bibfield  {journal} {\bibinfo
  {journal} {Physical Review B}\ }\textbf {\bibinfo {volume} {79}},\ \bibinfo
  {pages} {113403} (\bibinfo {year} {2009}{\natexlab{a}})}\BibitemShut
  {NoStop}%
\bibitem [{\citenamefont {Kang}\ \emph
  {et~al.}(2009{\natexlab{b}})\citenamefont {Kang}, \citenamefont {Shin},\ and\
  \citenamefont {Yeom}}]{Kang2009b}%
  \BibitemOpen
  \bibfield  {author} {\bibinfo {author} {\bibfnamefont {P.-G.}\ \bibnamefont
  {Kang}}, \bibinfo {author} {\bibfnamefont {J.~S.}\ \bibnamefont {Shin}}, \
  and\ \bibinfo {author} {\bibfnamefont {H.~W.}\ \bibnamefont {Yeom}},\ }\href
  {\doibase 10.1016/j.susc.2009.06.012} {\bibfield  {journal} {\bibinfo
  {journal} {Surf. Sci.}\ }\textbf {\bibinfo {volume} {603}},\ \bibinfo {pages}
  {2588} (\bibinfo {year} {2009}{\natexlab{b}})}\BibitemShut {NoStop}%
\bibitem [{\citenamefont {Nita}\ \emph {et~al.}(2011)\citenamefont {Nita},
  \citenamefont {Jalochowski}, \citenamefont {Krawiec},\ and\ \citenamefont
  {Stepniak}}]{Nita2011}%
  \BibitemOpen
  \bibfield  {author} {\bibinfo {author} {\bibfnamefont {P.}~\bibnamefont
  {Nita}}, \bibinfo {author} {\bibfnamefont {M.}~\bibnamefont {Jalochowski}},
  \bibinfo {author} {\bibfnamefont {M.}~\bibnamefont {Krawiec}}, \ and\
  \bibinfo {author} {\bibfnamefont {A.}~\bibnamefont {Stepniak}},\ }\href
  {\doibase 10.1103/PhysRevLett.107.026101} {\bibfield  {journal} {\bibinfo
  {journal} {Phys. Rev. Lett.}\ }\textbf {\bibinfo {volume} {107}},\ \bibinfo
  {pages} {026101} (\bibinfo {year} {2011})}\BibitemShut {NoStop}%
\bibitem [{\citenamefont {Krawiec}\ and\ \citenamefont
  {Jalochowski}(2013)}]{Krawiec2013}%
  \BibitemOpen
  \bibfield  {author} {\bibinfo {author} {\bibfnamefont {M.}~\bibnamefont
  {Krawiec}}\ and\ \bibinfo {author} {\bibfnamefont {M.}~\bibnamefont
  {Jalochowski}},\ }\href@noop {} {\bibfield  {journal} {\bibinfo  {journal}
  {Physical Review B}\ }\textbf {\bibinfo {volume} {87}},\ \bibinfo {pages}
  {5445} (\bibinfo {year} {2013})}\BibitemShut {NoStop}%
\bibitem [{\citenamefont {Ahn}\ \emph {et~al.}(2005)\citenamefont {Ahn},
  \citenamefont {Kang}, \citenamefont {Ryang},\ and\ \citenamefont
  {Yeom}}]{Ahn2005}%
  \BibitemOpen
  \bibfield  {author} {\bibinfo {author} {\bibfnamefont {J.~R.}\ \bibnamefont
  {Ahn}}, \bibinfo {author} {\bibfnamefont {P.~G.}\ \bibnamefont {Kang}},
  \bibinfo {author} {\bibfnamefont {K.~D.}\ \bibnamefont {Ryang}}, \ and\
  \bibinfo {author} {\bibfnamefont {H.~W.}\ \bibnamefont {Yeom}},\ }\href
  {\doibase 10.1103/PhysRevLett.95.196402} {\bibfield  {journal} {\bibinfo
  {journal} {Phys. Rev. Lett.}\ }\textbf {\bibinfo {volume} {95}},\ \bibinfo
  {pages} {196402} (\bibinfo {year} {2005})}\BibitemShut {NoStop}%
\bibitem [{\citenamefont {Snijders}\ \emph {et~al.}(2006)\citenamefont
  {Snijders}, \citenamefont {Rogge},\ and\ \citenamefont
  {Weitering}}]{Snijders2006}%
  \BibitemOpen
  \bibfield  {author} {\bibinfo {author} {\bibfnamefont {P.~C.}\ \bibnamefont
  {Snijders}}, \bibinfo {author} {\bibfnamefont {S.}~\bibnamefont {Rogge}}, \
  and\ \bibinfo {author} {\bibfnamefont {H.~H.}\ \bibnamefont {Weitering}},\
  }\href {\doibase 10.1103/PhysRevLett.96.076801} {\bibfield  {journal}
  {\bibinfo  {journal} {Phys. Rev. Lett.}\ }\textbf {\bibinfo {volume} {96}},\
  \bibinfo {pages} {076801} (\bibinfo {year} {2006})}\BibitemShut {NoStop}%
\bibitem [{\citenamefont {Snijders}\ and\ \citenamefont
  {Weitering}(2010)}]{Snijders2010}%
  \BibitemOpen
  \bibfield  {author} {\bibinfo {author} {\bibfnamefont {P.~C.}\ \bibnamefont
  {Snijders}}\ and\ \bibinfo {author} {\bibfnamefont {H.~H.}\ \bibnamefont
  {Weitering}},\ }\href {\doibase 10.1103/RevModPhys.82.307} {\bibfield
  {journal} {\bibinfo  {journal} {Reviews of Modern Physics}\ }\textbf
  {\bibinfo {volume} {82}},\ \bibinfo {pages} {307} (\bibinfo {year}
  {2010})}\BibitemShut {NoStop}%
\bibitem [{\citenamefont {Hasegawa}(2010)}]{Hasegawa2010}%
  \BibitemOpen
  \bibfield  {author} {\bibinfo {author} {\bibfnamefont {S.}~\bibnamefont
  {Hasegawa}},\ }\href {\doibase 10.1088/0953-8984/22/8/084026} {\bibfield
  {journal} {\bibinfo  {journal} {J. Phys.: Condens. Matter}\ }\textbf
  {\bibinfo {volume} {22}},\ \bibinfo {pages} {084026} (\bibinfo {year}
  {2010})}\BibitemShut {NoStop}%
\bibitem [{\citenamefont {Erwin}\ and\ \citenamefont
  {Himpsel}(2010)}]{Erwin2010}%
  \BibitemOpen
  \bibfield  {author} {\bibinfo {author} {\bibfnamefont {S.~C.}\ \bibnamefont
  {Erwin}}\ and\ \bibinfo {author} {\bibfnamefont {F.~J.}\ \bibnamefont
  {Himpsel}},\ }\href {\doibase 10.1038/ncomms1056} {\bibfield  {journal}
  {\bibinfo  {journal} {Nature Communications}\ }\textbf {\bibinfo {volume}
  {1}},\ \bibinfo {pages} {58} (\bibinfo {year} {2010})}\BibitemShut {NoStop}%
\bibitem [{\citenamefont {Biedermann}\ \emph {et~al.}(2012)\citenamefont
  {Biedermann}, \citenamefont {Regensburger}, \citenamefont {Fauster},
  \citenamefont {Himpsel},\ and\ \citenamefont {Erwin}}]{Biedermann2012}%
  \BibitemOpen
  \bibfield  {author} {\bibinfo {author} {\bibfnamefont {K.}~\bibnamefont
  {Biedermann}}, \bibinfo {author} {\bibfnamefont {S.}~\bibnamefont
  {Regensburger}}, \bibinfo {author} {\bibfnamefont {T.}~\bibnamefont
  {Fauster}}, \bibinfo {author} {\bibfnamefont {F.~J.}\ \bibnamefont
  {Himpsel}}, \ and\ \bibinfo {author} {\bibfnamefont {S.~C.}\ \bibnamefont
  {Erwin}},\ }\href {\doibase 10.1103/PhysRevB.85.245413} {\bibfield  {journal}
  {\bibinfo  {journal} {Physical Review B}\ }\textbf {\bibinfo {volume} {85}},\
  \bibinfo {pages} {245413} (\bibinfo {year} {2012})}\BibitemShut {NoStop}%
\bibitem [{\citenamefont {Snijders}\ \emph {et~al.}(2012)\citenamefont
  {Snijders}, \citenamefont {Johnson}, \citenamefont {Guisinger}, \citenamefont
  {Erwin},\ and\ \citenamefont {Himpsel}}]{Snijders2012}%
  \BibitemOpen
  \bibfield  {author} {\bibinfo {author} {\bibfnamefont {P.~C.}\ \bibnamefont
  {Snijders}}, \bibinfo {author} {\bibfnamefont {P.~S.}\ \bibnamefont
  {Johnson}}, \bibinfo {author} {\bibfnamefont {N.~P.}\ \bibnamefont
  {Guisinger}}, \bibinfo {author} {\bibfnamefont {S.~C.}\ \bibnamefont
  {Erwin}}, \ and\ \bibinfo {author} {\bibfnamefont {F.~J.}\ \bibnamefont
  {Himpsel}},\ }\href {\doibase 10.1088/1367-2630/14/10/103004} {\bibfield
  {journal} {\bibinfo  {journal} {New Journal of Physics}\ }\textbf {\bibinfo
  {volume} {14}},\ \bibinfo {pages} {103004} (\bibinfo {year}
  {2012})}\BibitemShut {NoStop}%
\bibitem [{\citenamefont {Kresse}\ and\ \citenamefont
  {Hafner}(1993)}]{Kresse1993}%
  \BibitemOpen
  \bibfield  {author} {\bibinfo {author} {\bibfnamefont {G.}~\bibnamefont
  {Kresse}}\ and\ \bibinfo {author} {\bibfnamefont {J.}~\bibnamefont
  {Hafner}},\ }\href@noop {} {\bibfield  {journal} {\bibinfo  {journal} {Phys.
  Rev. B}\ }\textbf {\bibinfo {volume} {47}},\ \bibinfo {pages} {558} (\bibinfo
  {year} {1993})}\BibitemShut {NoStop}%
\bibitem [{\citenamefont {Kresse}\ and\ \citenamefont
  {Furthm{\"u}ller}(1996)}]{Kresse1996}%
  \BibitemOpen
  \bibfield  {author} {\bibinfo {author} {\bibfnamefont {G.}~\bibnamefont
  {Kresse}}\ and\ \bibinfo {author} {\bibfnamefont {J.}~\bibnamefont
  {Furthm{\"u}ller}},\ }\href@noop {} {\bibfield  {journal} {\bibinfo
  {journal} {Phys. Rev. B}\ }\textbf {\bibinfo {volume} {54}},\ \bibinfo
  {pages} {11169} (\bibinfo {year} {1996})}\BibitemShut {NoStop}%
\bibitem [{\citenamefont {Blochl}(1994)}]{Blochl1994}%
  \BibitemOpen
  \bibfield  {author} {\bibinfo {author} {\bibfnamefont {P.~E.}\ \bibnamefont
  {Blochl}},\ }\href@noop {} {\bibfield  {journal} {\bibinfo  {journal} {Phys.
  Rev. B}\ }\textbf {\bibinfo {volume} {50}},\ \bibinfo {pages} {17953}
  (\bibinfo {year} {1994})}\BibitemShut {NoStop}%
\bibitem [{\citenamefont {Kresse}\ and\ \citenamefont
  {Joubert}(1999)}]{Kresse1999}%
  \BibitemOpen
  \bibfield  {author} {\bibinfo {author} {\bibfnamefont {G.}~\bibnamefont
  {Kresse}}\ and\ \bibinfo {author} {\bibfnamefont {D.}~\bibnamefont
  {Joubert}},\ }\href {\doibase 10.1103/PhysRevB.59.1758} {\bibfield  {journal}
  {\bibinfo  {journal} {Physical Review B}\ }\textbf {\bibinfo {volume} {59}},\
  \bibinfo {pages} {1758} (\bibinfo {year} {1999})}\BibitemShut {NoStop}%
\bibitem [{\citenamefont {Shin}\ \emph {et~al.}(2012)\citenamefont {Shin},
  \citenamefont {Ryang},\ and\ \citenamefont {Yeom}}]{Shin2012}%
  \BibitemOpen
  \bibfield  {author} {\bibinfo {author} {\bibfnamefont {J.~S.}\ \bibnamefont
  {Shin}}, \bibinfo {author} {\bibfnamefont {K.-D.}\ \bibnamefont {Ryang}}, \
  and\ \bibinfo {author} {\bibfnamefont {H.~W.}\ \bibnamefont {Yeom}},\ }\href
  {\doibase 10.1103/PhysRevB.85.073401} {\bibfield  {journal} {\bibinfo
  {journal} {Physical Review B}\ }\textbf {\bibinfo {volume} {85}},\ \bibinfo
  {pages} {073401} (\bibinfo {year} {2012})}\BibitemShut {NoStop}%
\bibitem [{\citenamefont {Rangarajan}\ and\ \citenamefont
  {Ding}(2000)}]{Rangarajan2000}%
  \BibitemOpen
  \bibfield  {author} {\bibinfo {author} {\bibfnamefont {G.}~\bibnamefont
  {Rangarajan}}\ and\ \bibinfo {author} {\bibfnamefont {M.~Z.}\ \bibnamefont
  {Ding}},\ }\href {\doibase 10.1103/PhysRevE.61.4991} {\bibfield  {journal}
  {\bibinfo  {journal} {Physical Review E}\ }\textbf {\bibinfo {volume} {61}},\
  \bibinfo {pages} {4991} (\bibinfo {year} {2000})}\BibitemShut {NoStop}%
\bibitem [{\citenamefont {Mermin}\ and\ \citenamefont
  {Wagner}(1966)}]{Mermin1966}%
  \BibitemOpen
  \bibfield  {author} {\bibinfo {author} {\bibfnamefont {N.~D.}\ \bibnamefont
  {Mermin}}\ and\ \bibinfo {author} {\bibfnamefont {H.}~\bibnamefont
  {Wagner}},\ }\href {\doibase 10.1103/PhysRevLett.17.1133} {\bibfield
  {journal} {\bibinfo  {journal} {Physical Review Letters}\ }\textbf {\bibinfo
  {volume} {17}},\ \bibinfo {pages} {1133} (\bibinfo {year}
  {1966})}\BibitemShut {NoStop}%
\bibitem [{\citenamefont {Erwin}\ and\ \citenamefont
  {Hellberg}(2005)}]{Erwin2005a}%
  \BibitemOpen
  \bibfield  {author} {\bibinfo {author} {\bibfnamefont {S.~C.}\ \bibnamefont
  {Erwin}}\ and\ \bibinfo {author} {\bibfnamefont {C.~S.}\ \bibnamefont
  {Hellberg}},\ }\href {\doibase 10.1016/j.susc.2005.04.026} {\bibfield
  {journal} {\bibinfo  {journal} {Surface Science}\ }\textbf {\bibinfo {volume}
  {585}},\ \bibinfo {pages} {L171} (\bibinfo {year} {2005})}\BibitemShut
  {NoStop}%
\bibitem [{\citenamefont {Bruch}(2005)}]{Bruch2005}%
  \BibitemOpen
  \bibfield  {author} {\bibinfo {author} {\bibfnamefont {L.~W.}\ \bibnamefont
  {Bruch}},\ }\href {\doibase 10.1016/j.susc.2005.04.055} {\bibfield  {journal}
  {\bibinfo  {journal} {Surface Science}\ }\textbf {\bibinfo {volume} {585}},\
  \bibinfo {pages} {135} (\bibinfo {year} {2005})}\BibitemShut {NoStop}%
\end{thebibliography}
%merlin.mbs apsrev4-1.bst 2010-07-25 4.21a (PWD, AO, DPC) hacked
%Control: key (0)
%Control: author (8) initials jnrlst
%Control: editor formatted (1) identically to author
%Control: production of article title (-1) disabled
%Control: page (0) single
%Control: year (1) truncated
%Control: production of eprint (0) enabled
%

\end{document}